\documentclass[%
 aip,
 amsmath,amssymb,
 reprint,%
]{revtex4-1}

\usepackage[pdftex]{graphicx}
\usepackage{dcolumn}
\usepackage{bm}

\usepackage[utf8]{inputenc}
\usepackage[T1]{fontenc}
\usepackage{mathptmx}
\usepackage{etoolbox}

\usepackage{hyperref}

\makeatletter
\def\@email#1#2{%
 \endgroup
 \patchcmd{\titleblock@produce}
  {\frontmatter@RRAPformat}
  {\frontmatter@RRAPformat{\produce@RRAP{*#1\href{mailto:#2}{#2}}}\frontmatter@RRAPformat}
  {}{}
}%
\makeatother
\begin{document}

\preprint{AIP/123-QED}

\title{PyAPX: Python toolkit for atomic configuration pattern exploration}

\author{A. Kusaba}
 \email{kusaba@riam.kyushu-u.ac.jp}
\affiliation{%
 Research Institute for Applied Mechanics, Kyushu University, Fukuoka 816-8580, Japan
}%
\affiliation{%
Institute of Materials and Systems for Sustainability, Nagoya University, Nagoya 464-8601, Japan
}%

\author{T. Kuboyama}
\affiliation{
Computer Centre, Gakushuin University, Toshima-ku, Tokyo 171-8588, Japan
}

\author{K. Kawka}
\author{P. Kempisty}%
\affiliation{%
Institute of High Pressure Physics, Polish Academy of Sciences, Sokolowska 29/37, 01-142 Warsaw, Poland
}%

\author{Y. Kangawa}
\affiliation{%
 Research Institute for Applied Mechanics, Kyushu University, Fukuoka 816-8580, Japan
}%

\date{\today}

\begin{abstract}
In materials discovery, the integration of first-principles calculations with machine learning techniques has been actively studied for two key tasks: crystal structure prediction, which searches for stable structures given a chemical composition, and elemental substitution, which explores chemical compositions that yield desirable properties in a given crystal structure. 
However, even when both the crystal structure and chemical composition are fixed, material properties can still vary depending on the atomic arrangements (configurations) at crystallographic sites. 
To support detailed material design, we present PyAPX, a Python toolkit that performs Bayesian searches of stable atomic configurations. 
A distinctive feature of this initial release is the introduction of encoding methods suitable for configuration search, and we evaluate their performance using the h-BCN system.
As a result, they were confirmed to yield superior convergence compared to commonly used one-hot encoding. 
PyAPX is broadly applicable to crystalline materials and is expected to further advance materials discovery.
\end{abstract}

\maketitle

\section{Introduction}
Materials discovery approaches based on first-principles calculations have been studied in the field of crystal structure prediction (CSP). 
The task of CSP is to obtain stable crystal structures given a chemical composition as input (Fig.~1(a)). 
Representative software includes USPEX~\cite{oganov2006crystal,oganov2011evolutionary,lyakhov2013new}, which is based on evolutionary algorithms, and CALYPSO~\cite{wang2010crystal,wang2012calypso}, which employs particle swarm optimization. 
One notable achievement of CSP is a series of computational predictions aiming at room-temperature superconductivity~\cite{wang2012superconductive,peng2017hydrogen,liu2017potential,kruglov2018superconductivity}. 
Subsequently, superconductivity at 250~K in LaH$_{10}$ was experimentally reported~\cite{drozdov2019superconductivity}.

\begin{figure}
\includegraphics[width=8.5cm]{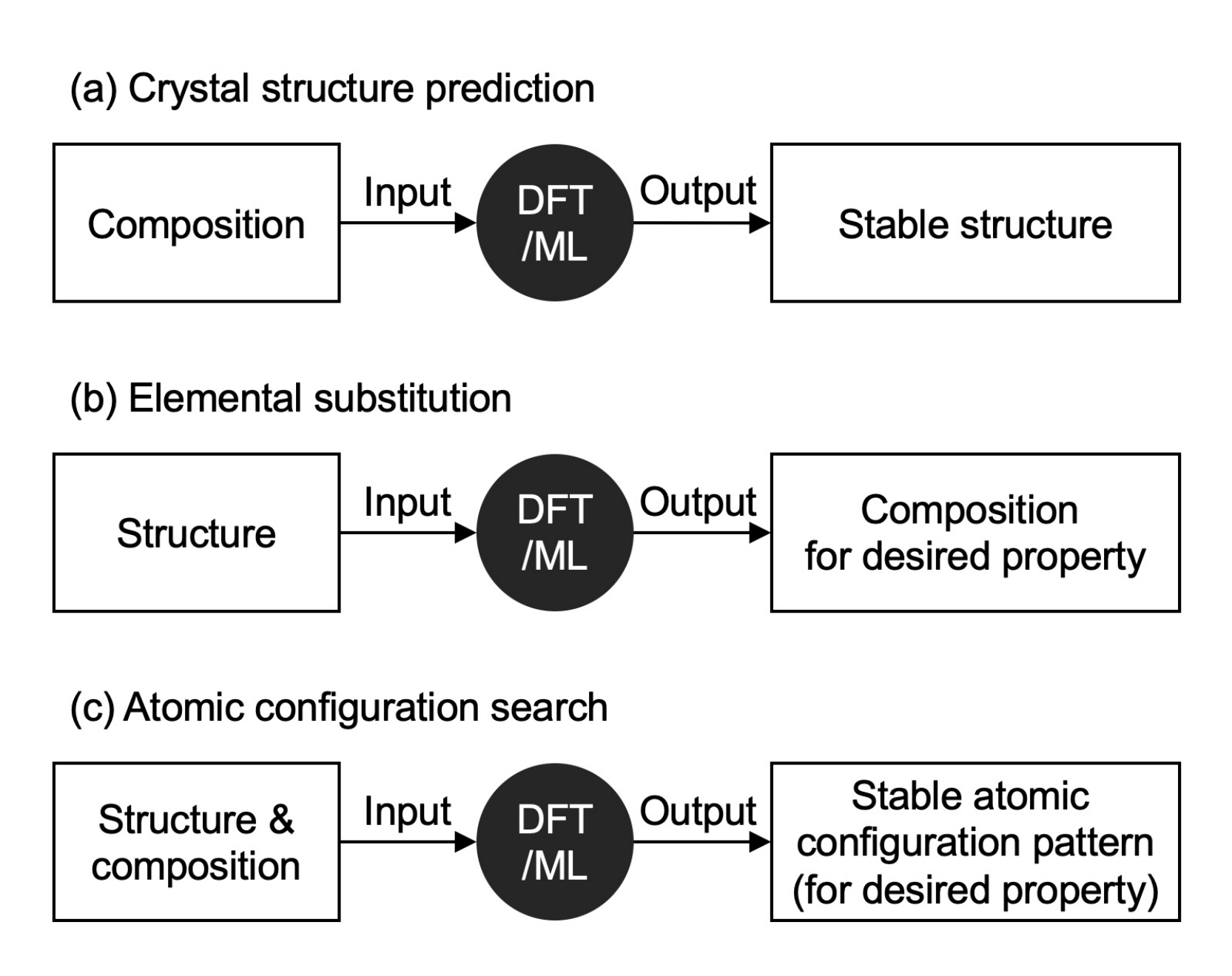}
\caption{(a, b) Typical problem settings in materials discovery based on first-principles calculations (and machine learning), and (c) the problem setting focused on in this study. Input refers to the design variables that are fixed in the discovery system, while Output denotes those to be optimized.}
\end{figure}

Another materials discovery strategy based on first-principles calculations is elemental substitution~\cite{nishijima2014accelerated,iwasaki2022efficient}. 
In this strategy, a base crystal structure is given, and various elements from the periodic table are substituted, followed by high-throughput calculations to identify materials with desirable properties in the vast composition space (Fig.~1(b)). 
For example, a cathode material with improved cycle life for lithium-ion batteries was discovered through about 2000 density functional theory (DFT) calculations for different elemental substitutions of LiFePO$_4$~\cite{nishijima2014accelerated}.

More recently, materials discovery methods incorporating machine learning techniques have been actively developed~\cite{iwasaki2022efficient,du2023machine,han2025efficient}. 
For example, Microsoft’s MatterGen~\cite{zeni2025generative} unifies the tasks of CSP and elemental substitution by jointly generating composition and structure. 
It is a diffusion-based generative model that enables the direct generation (computational design) of new materials given desired property constraints.
Also, Google DeepMind’s GNoME~\cite{merchant2023scaling} explores crystal structures and compositions by predicting their stability using graph neural networks. 

Even after both composition and structure (more precisely, the lattice and atomic coordinates) are determined, multicomponent systems still retain degrees of freedom in the atomic arrangements (configurations) at crystallographic sites. 
In the field of solid-solution materials, where the aim is often to mix multiple components to create new materials, methods such as the special quasirandom structure (SQS) approach~\cite{zunger1990special,van2009multicomponent,van2013efficient} have been applied to sample well-mixed atomic configurations.
In addition, in recent years, sampling methods for statistically treating the intermediate levels of disorder have also been advanced by incorporating machine learning techniques~\cite{kasamatsu2023configuration}.

On the other hand, it is also worth considering the task of searching for specific, non-random stable atomic configuration patterns (Fig.~1(c)), since atomic configurations can significantly affect material properties. 
For example, in the two-dimensional material h-BCN, the band gap is known to vary considerably depending on the atomic configuration, even at a fixed composition~\cite{liu1989atomic,azevedo2006structural,zhu2011interpolation,hara2025exploration}. 
Also, searching for adsorption configurations on crystal surfaces~\cite{kusaba2022exploration,kawka2024augmentation,kempisty2023polar,kuboyama2025sat} falls within the scope of this task. 
Although not a materials discovery task, such studies are essential for identifying surface reconstructions and designing crystal growth processes.

Exploration of the degrees of freedom in atomic configurations can be regarded as being encompassed within the CSP framework. 
However, it can be addressed more directly as a combinatorial optimization problem, in which the continuous degrees of freedom of the lattice and atomic coordinates are frozen (as initial coordinates before relaxation), and atoms are rearranged under a given compositional constraint to achieve stabilization. 
In fact, successful cases have been reported in which approaches based on Bayesian optimization have solved the combinatorial optimization of atomic configurations~\cite{kusaba2022exploration,kawka2024augmentation,hara2025exploration,ono2022optimization}.

Overall, atomic configuration pattern exploration has received relatively little attention so far, but it represents an important problem setting for achieving more detailed materials design and advancing materials discovery to the next stage. 
However, user-friendly tools enabling stable atomic configuration searches by coupling Bayesian optimization libraries with first-principles codes remain underdeveloped. 
To address this, we present PyAPX as a practical solution.

\section{Methods}
\subsection{Scheme of the PyAPX Workflow}
The PyAPX workflow is organized into a pre-process that prepares the candidate pool and a main process that sequentially repeats sampling and energy evaluation (Fig.~2).
In the pre-process, the user provides a list of atomic configurations, thereby defining the search space, which is then transformed into feature vectors.
The loop of the main process continues until the specified number of sampling iterations is reached, while the sampled atomic configurations and their corresponding DFT total energies are recorded during the process.

Sampling from the candidate pool is conducted within the framework of Bayesian optimization in order to identify more stable atomic configurations.
At each iteration, a Bayesian model predicts the DFT total energies of the candidate structures, together with their uncertainties, based on the data acquired so far.
Subsequently, all candidate structures are typically scored using an acquisition function based on those model outputs, and the structure to be sampled is determined.
In Bayesian optimization, the sampling strategy does not simply select the candidate with the best predicted value, but also explores regions of the domain where data are sparse and the prediction uncertainty is high.
By balancing exploration and exploitation in this way, efficient global optimization can be achieved.
At present, PyAPX relies on the Bayesian optimization implementation provided by the Python library PHYSBO~\cite{motoyama2022bayesian,ueno2016combo}.

Next, the sampled structure is passed to the energy evaluator.
Typically, a DFT code serves as the energy evaluator: the atomic-coordinate section of a template for the DFT input file prepared by the user is modified, a DFT calculation is executed, and the total energy of the sampled structure is obtained.
In this release, only Quantum ESPRESSO~\cite{giannozzi2017advanced,giannozzi2009quantum} is supported, but support for other commonly used DFT codes is planned for future versions.
Alternatively, a user-defined function that takes atomic configurations as input and outputs energy can also be used as an energy evaluator.
This option enables users to interface PyAPX with model calculations such as the Ising model, as well as with unsupported DFT codes or machine-learning interatomic-potential programs.

\begin{figure}
\includegraphics[width=8.5cm]{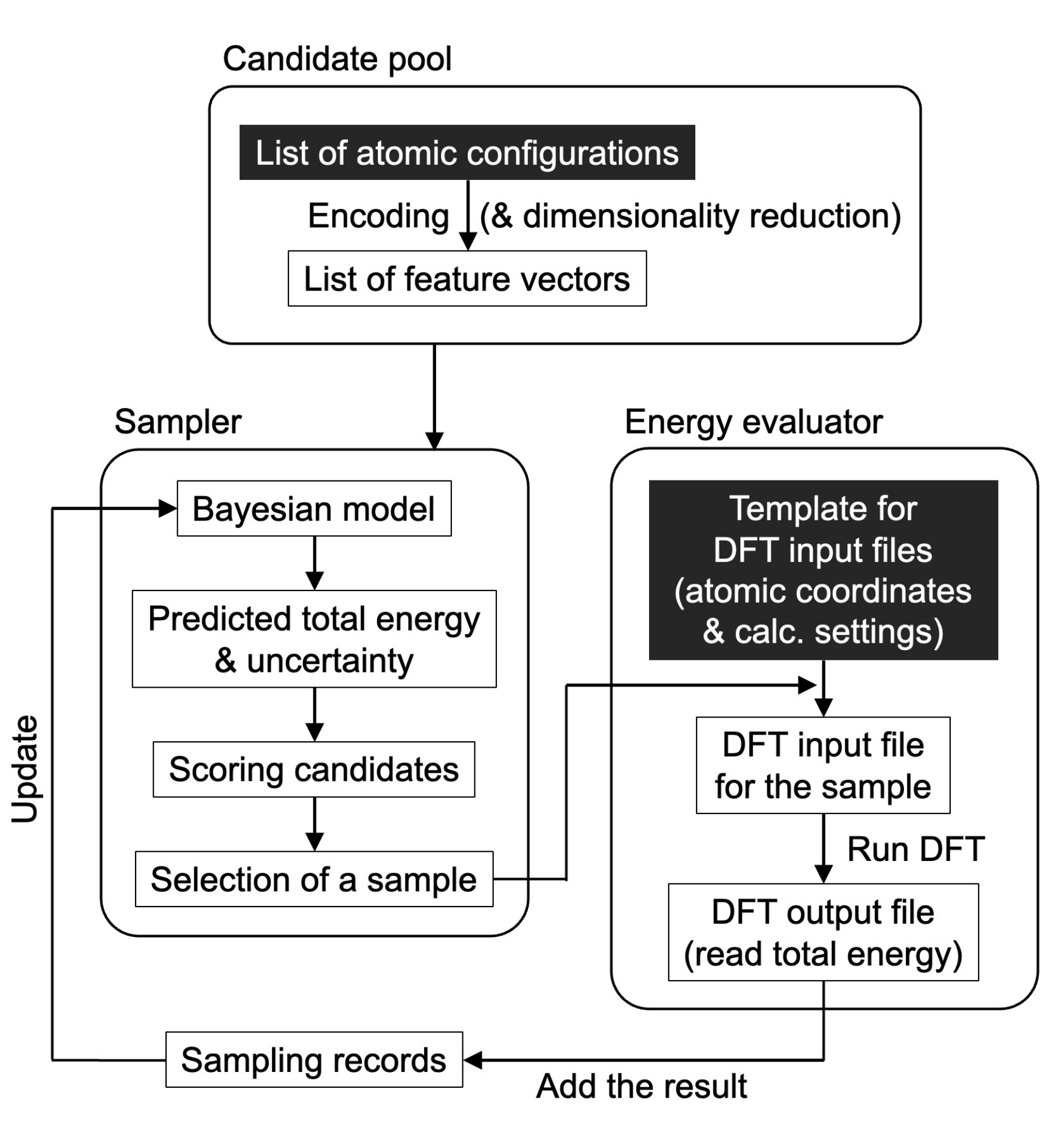}
\caption{Scheme of the PyAPX workflow. Items in black boxes are prepared by the user. The DFT input files are automatically generated within the sequential loop of sampling and DFT calculations, which continues until the specified number of sampling iterations is reached.}
\end{figure}

\subsection{Encoding Methods for Atomic Configurations}
To efficiently perform Gaussian-kernel-based Bayesian optimization, it is essential to effectively embed each candidate pattern of atomic configuration into the input space of the Bayesian model.
One effective prescription for this is to incorporate sufficient information about the local atomic environments when representing each candidate as a vector.
Here, using a material consisting of three elements, Z1, Z2, and Z3, as an example, we introduce two additional encoding options, namely the neighbor-atom (NA) encoding and the modified neighbor-atom (NAmod) encoding, in addition to the commonly used one-hot encoding~\cite{kusaba2022exploration,ono2022optimization}. 
These two encoding options contain richer information on the local atomic environments and can be applied to a wide range of crystal structures, not limited to the honeycomb lattice example shown in Fig.~3.

\begin{figure}
\includegraphics[width=8.5cm]{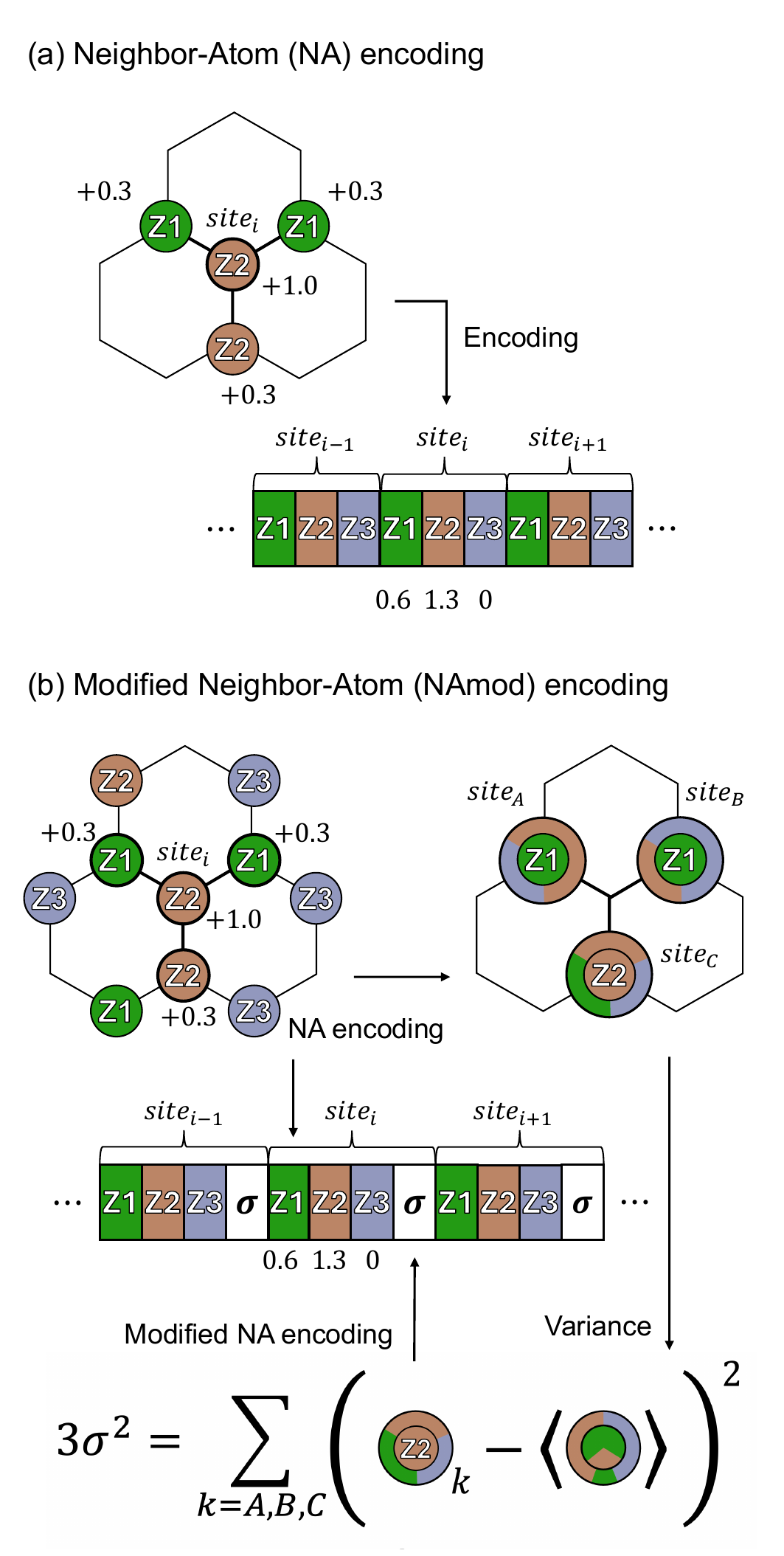}
\caption{Schematic illustration of (a) the neighbor-atom (NA) encoding and (b) the modified neighbor-atom (NAmod) encoding, using a honeycomb lattice system composed of three elements, Z1, Z2, and Z3, as an example. Green, brown, and purple circles represent Z1, Z2, and Z3 atoms, respectively, and the multicolored double circles indicate the convolution of atomic occupancies at neighboring sites.}
\end{figure}

First, indices are assigned to the crystallographic sites where atoms are located.
In the one-hot encoding scheme, each \textit{site-i} has values ($x_{i\text{-}Z1}$, $x_{i\text{-}Z2}$, $x_{i\text{-}Z3}$) corresponding to the three considered elements in the present case.
When an element Z ($=$ Z1, Z2 or Z3) is assigned to \textit{site-i}, the value $x_{i\text{-}Z}$ corresponding to that element is set to 1, and the remaining values are set to 0, representing the atomic occupancy of \mbox{\textit{site-i}}.
Thus, a candidate structure is represented by a vector with dimensions equal to the number of sites multiplied by the number of elements.

However, in the one-hot encoding, information about the spatial relationships among sites is not utilized in the optimization process.
In the NA encoding, the one-hot encoding is modified to incorporate information about the atomic occupancy of neighboring sites, and the feature vector $\boldsymbol{\phi}^{\text{NA}}_i$ is expressed as follows:
\begin{equation}
\boldsymbol{\phi}^{\text{NA}}_i = 
\begin{pmatrix}
x_{i\text{-}Z1} \\
x_{i\text{-}Z2} \\
x_{i\text{-}Z3}
\end{pmatrix}
+
w
\begin{pmatrix}
n_{i\text{-}Z1} \\
n_{i\text{-}Z2} \\
n_{i\text{-}Z3}
\end{pmatrix}
\end{equation}
Here, the first term on the right-hand side corresponds to the \textit{site-i} components of the one-hot encoding, while the second term adds information from the neighboring sites.
Specifically, the weight $w$ represents the degree of information incorporation from the neighboring sites, and the neighboring atom counts $n_{i\text{-}Z}$ indicate the number of atoms of element Z located among the neighboring sites of \textit{site-i} (three neighboring sites for this honeycomb lattice case).
An illustrative example for $w=0.3$ is shown in Fig.~3(a).
This formulation can be interpreted as a convolution of the atomic occupancy of neighboring sites.

Furthermore, to incorporate information on the anisotropy in the local atomic environments, we propose a modified NA encoding.
As illustrated in Fig.~3(b), the local anisotropy around \textit{site-i} is formulated as the variance $\sigma_i^2$ of the components of the NA encoding vector corresponding to each neighboring site (\textit{site-A}, \textit{site-B}, and \textit{site-C} in this case).
\begin{equation}
\sigma_i^2 = \sum_k d(\boldsymbol{\phi}^{\text{NA}}_k, \sum_k \boldsymbol{\phi}^{\text{NA}}_k / N^{neighbor}_i)^2 / N^{neighbor}_i
\end{equation}
Here, $d(\bullet, \bullet)$ denotes the Euclidean distance, $N^{neighbor}_i$ is the number of neighboring sites of \textit{site-i}, and $k$ represents the index of each neighboring site of \textit{site-i}.
The standard deviation $\sigma_i$ is then added as a descriptor for \textit{site-i}.
\begin{equation}
\boldsymbol{\phi}^{\text{NAmod}}_i = 
\begin{pmatrix}
\boldsymbol{\phi}^{\text{NA}}_i \\
\sigma_i
\end{pmatrix}
\end{equation}
Consequently, in the modified NA encoding, each candidate structure is represented by a vector whose dimension is greater by the number of sites than that of the one-hot and NA encodings.

\subsection{Demonstration Setup: h‑BCN System}
Hexagonal boron carbon nitride (h-BCN) is a two-dimensional honeycomb-lattice material that can be regarded as an alloy of graphene and h-BN. 
It has been reported that, even when the chemical composition is fixed, the band gap varies significantly depending on the atomic configuration pattern~\cite{liu1989atomic,azevedo2006structural,zhu2011interpolation}.
In this study, we adopted a (3$\times$3) periodic \mbox{h-BCN} system (Fig.~4) for demonstration purposes, as its DFT computational cost is relatively low.
This system contains 18 sites, which are occupied by six B atoms, six C atoms, and six N atoms. 
Because all possible atomic configurations can be translationally shifted so that a C atom is located at \mbox{\textit{site-}1}, fixing a C atom at \textit{site-}1 does not reduce the generality of the system.
When other symmetries are not considered for simplicity, the total number of possible configurations is ${}_{17}C_{6} \times {}_{11}C_{6} \times {}_{5}C_{5}$ = 5717712, which provides a configuration space of a suitable scale for demonstration.

\begin{figure}
\includegraphics[width=8.5cm]{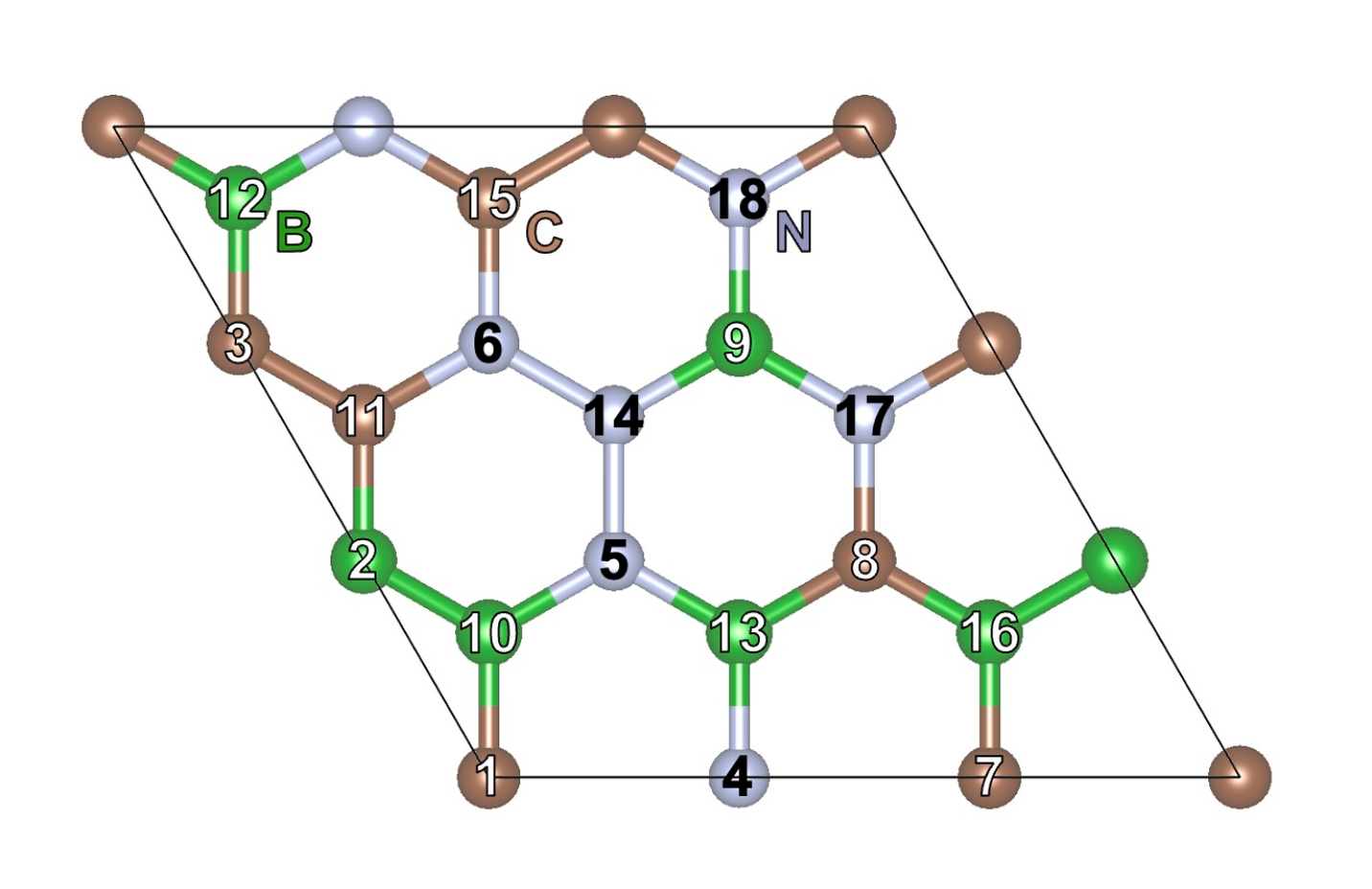}
\caption{An example of an atomic configuration pattern in the (3$\times$3) periodic h-BCN system. This system contains 18 sites, and the numbers in the figure indicate their site indices. Six B atoms, six C atoms, and six N atoms, represented by green, brown, and blue spheres, respectively, occupy the sites. The crystal structure was visualized using the VESTA~\cite{momma2011vesta}.}
\end{figure}

In this study, we compare the exploration performance obtained using the one-hot, NA, and modified NA encodings.
In the present case, these methods yield feature vectors of 54, 54, and 72 dimensions, respectively.
In addition, we also consider a case in which the feature vectors obtained by the modified NA encoding are reduced to 54 dimensions by principal component analysis (PCA), so that the resulting dimensionality matches that of the one-hot and NA encodings; this case is referred to as NAmod+PCA.
For the three cases other than one-hot, the parameter $w$ in the encoding was set to 0.3, so that the maximum possible contribution from the neighboring sites, $3w$, would not exceed the central-site occupation value of 1.

The sampling conditions were set as follows. 
First, initial data were collected through 100 random samplings and corresponding DFT calculations. 
The same initial dataset was used for all four encoding cases. 
Subsequently, Bayesian sampling and DFT calculations were sequentially performed 400 times for each encoding case, aiming to minimize the DFT total energy of the system obtained after structural relaxation.
In PHYSBO, random feature mapping~\cite{rahimi2007random} and Thompson sampling~\cite{chapelle2011empirical} are implemented to accelerate Bayesian optimization by reducing the computational cost of the acquisition function. 
In this study, these functionalities were utilized, and the dimensionality of the random feature map was set to 3000. 
In addition, the hyperparameters in the Bayesian optimization were updated at each sampling step based on the maximization of the type-II likelihood~\cite{rasmussen2003gaussian}. 
For further details on these functionalities, please refer to the PHYSBO documentation~\cite{motoyama2022bayesian}.

The DFT calculations were performed using Quantum ESPRESSO~\cite{giannozzi2017advanced,giannozzi2009quantum}. 
The exchange-correlation energy was treated within the generalized gradient approximation (GGA) using the Perdew-Burke-Ernzerhof (PBE) functional~\cite{perdew1996generalized}. 
Projector-augmented wave (PAW)~\cite{kresse1999ultrasoft} pseudopotentials were employed.
The cutoff energies for the wavefunctions and the charge densities were set to 60 and 240 Ry, respectively.
The k-point sampling was carried out with a 3$\times$3$\times$1 grid. 
Structural optimizations were conducted with convergence thresholds of $10^{-5}$ Ry for energy and $10^{-4}$ Ry/Bohr for forces. 
In addition, the cell parameters were also optimized for each sampled structure during the structural relaxation process.

\section{Results and Discussion}
Figures 5(a)--(d) show the sampling histories obtained by Bayesian optimization using the one-hot, NA, modified NA, and modified NA with PCA encodings, respectively.
The vertical axis represents the DFT total energy after structural relaxation, which is the target quantity to be minimized during the Bayesian sampling process.
It is plotted as a relative value with respect to the composition-weighted average of the total energies of the pure graphene and h-BN phases.
Thus, the vertical axis corresponds to the so-called mixing enthalpy, $\Delta H_{mix}$.
In these panels, to visualize the sampling behavior more clearly, statistical quantities are also shown: the cumulative minimum (black solid line), moving average (red solid line), and moving standard deviation (purple shaded area).
The window size for the latter two moving statistical quantities was set to 51 samples.
Figures 5(e)--(h) show plots of these statistical quantities to facilitate comparison among the different encoding methods.

\begin{figure*}
\includegraphics[width=\textwidth]{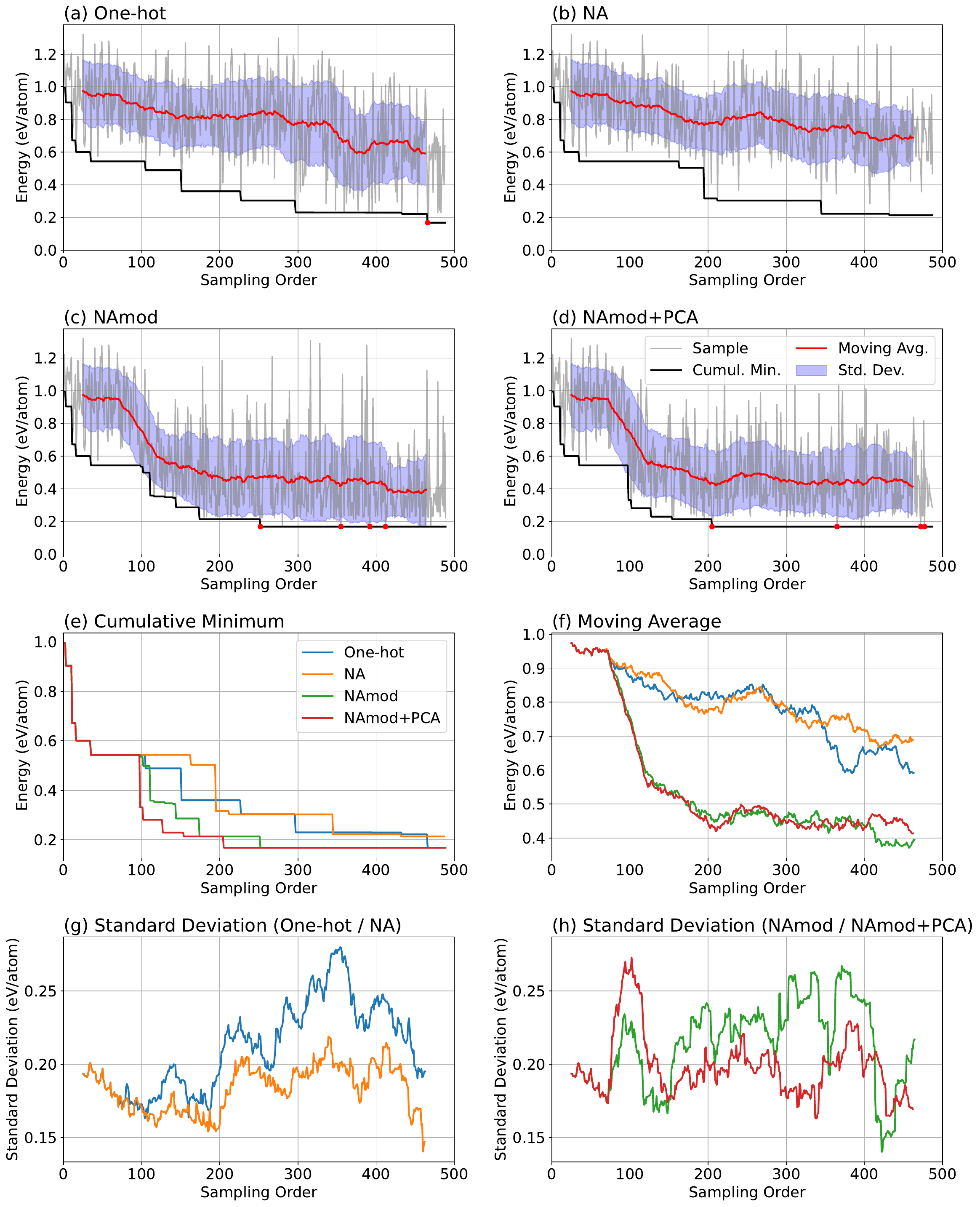}
\caption{Sampling histories obtained by Bayesian optimization for each encoding case: (a) one-hot, (b) NA, (c) modified NA, and (d) modified NA with PCA. The legends are shared among panels (a)--(d), and the red circle markers correspond to the most stable configurations. Statistical plots for comparing their optimization performance: (e) cumulative minimum, (f) moving average, and (g)--(h) moving standard deviation. The legends are shared among panels (e)--(h), and the window size for the moving averages and moving standard deviations was set to 51 samples.}
\end{figure*}

First, even with the simplest one-hot encoding, a decrease in the moving average can be observed, confirming the effectiveness of the Bayesian optimization.
One of the most stable configurations, at least within the range recognized across the four sampling series in this study, was also identified (indicated by red circle markers in the figure).
In the case of the NA encoding, no improvement in optimization performance is observed compared with the one-hot case in terms of the cumulative minimum or moving average.
This suggests that a simple convolution of elemental information from neighboring sites is insufficient to adequately characterize the configuration patterns in this system.
Nevertheless, the NA encoding does exhibit a certain effect, which appears as a reduction in the moving standard deviation.

In the modified NA encoding, which incorporates information on the anisotropy of local atomic environments, a clear improvement in performance was observed in both the cumulative minimum and moving average compared with the two aforementioned encoding methods.
This indicates that such local anisotropy constitutes an important feature of this material system.
Furthermore, four symmetry-equivalent most stable configurations were successfully identified.
In the case with dimensionality reduction using PCA, the moving average was comparable to that of the case without PCA, while a faster decrease in the cumulative minimum was observed.
In general, in the one-hot encoding (and in the NA encoding based on it), the occupation of the three elements (B, C, or N) at a given site can be represented in two dimensions, but in the present implementation it is represented in three dimensions.
In this demonstration, the dimensionality reduction by PCA was limited to eliminating this redundant dimension.
Indeed, the cumulative explained variance ratio of the retained components reached 1.000, indicating that the reduced representation preserves essentially all of the variance present in the original (NAmod) feature space.
Therefore, the observed improvement is considered to reflect that the training of the Bayesian model and the optimization of atomic configurations were facilitated by the reduced input dimensionality, rather than a degradation of optimization performance caused by loss of encoding information.

However, the above results are influenced by stochastic factors, including random sampling of the initial data, random feature mapping, and Thompson sampling. To obtain a more reliable performance evaluation, we conducted four additional comparative experiments following the same protocol (five independent experiments in total). The initial datasets were independently resampled for each of the additional experiments, while being shared across the different encoding methods within each experiment, as in the first experiment. The results of all five independent experiments are summarized in Fig.~6. Here, the solid lines represent the average cumulative minimum over the five experiments, and the shaded area indicates the range defined by the maximum and minimum cumulative minima across the experiments.
On average, the NA encoding exhibited rather worse performance than the one-hot encoding. In contrast, both NAmod and NAmod+PCA consistently showed faster optimization than the one-hot encoding, confirming the observed trend on average.

\begin{figure}
\includegraphics[width=\linewidth]{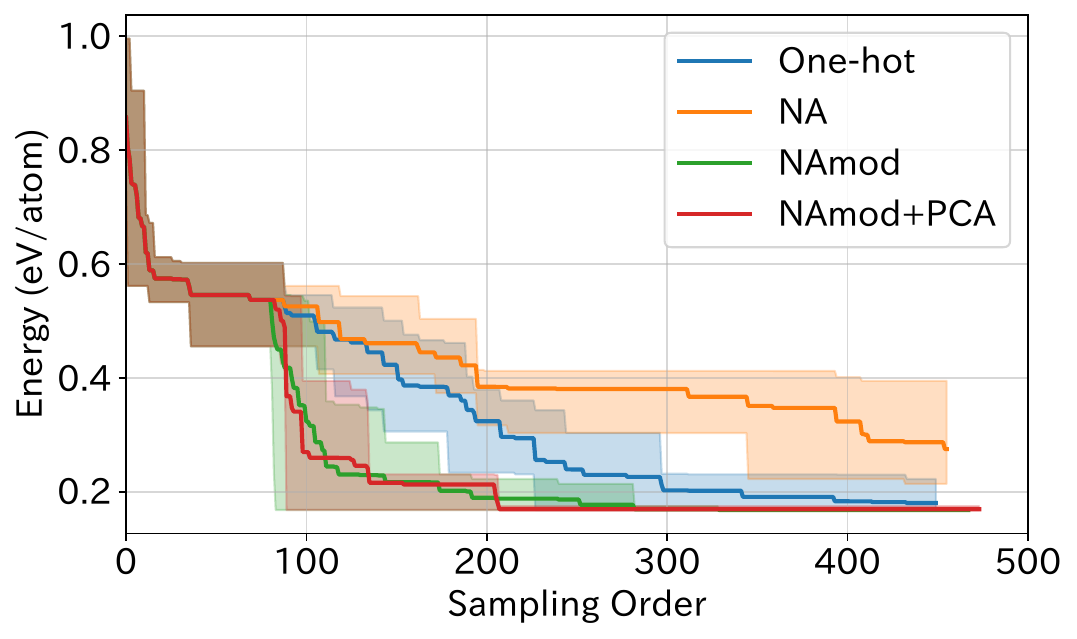}
\caption{Average cumulative minimum over five experiments (solid line), with the shaded area representing the range of cumulative minima across the experiments.}
\end{figure}

\section{Conclusions}
In this study, we presented PyAPX, a user-friendly Python toolkit that enables stable atomic configuration searches by coupling a Bayesian optimization library with DFT codes.
We also proposed several encoding methods for configuration exploration and compared their performance.
The commonly used one-hot encoding, which is also implemented in PyAPX, can serve as a reasonable first choice for new material systems, as it is easy to set up while being reasonably functional.
However, the proposed NAmod (modified neighbor-atom) encoding demonstrated higher optimization performance in the h-BCN system.
This encoding can therefore be a promising option for material systems in which the stable configuration search is difficult using one-hot encoding.
Furthermore, PyAPX, which is designed to be widely applicable to crystalline materials, is expected to advance materials discovery to the next stage.

\begin{acknowledgments}
We thank David Bowler for helpful discussions.
This work was partially supported by JSPS KAKENHI (grant numbers JP23K28151, JP24K17619, JP24H00432); JST BOOST (grant number JPMJBY24C3); and Collaborative Research Program of Research Institute for Applied Mechanics, Kyushu University. 
The computation was carried out using the computer resource offered under the category of General Projects by Research Institute for Information Technology, Kyushu University; and using Research Center for Computational Science, Okazaki, Japan (Project: 25-IMS-C129).
\end{acknowledgments}

\section*{Data Availability}
The data that support the findings of this study are available from the corresponding author upon reasonable request. 
In addition, the PyAPX toolkit is publicly available on GitHub at \url{https://github.com/a-ksb/PyAPX}.

%

\end{document}